\input{aipcheck}

\documentclass[
    ,final            
  ]
  {aipproc}

\layoutstyle{8x11single}

\def\A {\mathcal{A}}

\def\DKpp{D^+  \to K^- \pi^+\pi^+ }

\begin{document}

\title{$K\pi$ form factors, final state interactions and $D^+\to K^-\pi^+\pi^+$ decays\footnote{
Talk given by R.~Escribano
at HADRON 2009: XIII International Conference on Hadron Spectroscopy,
Florida, USA, 11/29-12/04/2009}}


\author{D.~R.~Boito and R.~Escribano}{
  address={Grup de F\'{\i}sica Te\`orica and IFAE, Universitat Aut\`onoma de Barcelona,\\
                    E-08193 Bellaterra (Barcelona), Spain}
}

\begin{abstract}
We present a model for the decay $D^+\to K^-\pi^+\pi^+$. 
The weak interaction part of this reaction is described using the effective weak Hamiltonian
in the factorisation approach.
Hadronic final state interactions are taken into account through the $K\pi$ scalar and vector
form factors fulfilling analyticity, unitarity and chiral symmetry constraints.
Allowing for a global phase difference between the $S$ and $P$ waves of $-65^\circ$,
the Dalitz plot of the $D^+\to K^-\pi^+\pi^+$ decay, the $K\pi$ invariant mass spectra and
the total branching ratio due to $S$-wave interactions are well reproduced.
\end{abstract}

\maketitle


\section{Introduction}

In 2002, the analysis of $\DKpp$ decays performed by the E791 collaboration revealed that approximately 50\% of these decays proceed through a low-mass scalar resonance with isospin $1/2$: the $K_0^*(800)$, also called the $\kappa$~\cite{Aitala:2002kr}. 
More recently, the $\DKpp$ decay was revisited by E791~\cite{Aitala:2005yh} and two other experiments produced analyses based on larger data samples,
namely FOCUS~\cite{Pennington:2007se,Focus2009} and CLEO~\cite{Bonvicini:2008jw}.
The main conclusions of the pioneering E791 work have been confirmed in both cases.

In the past, many analyses of $K\pi$ scattering data had already claimed the presence of the
$\kappa$ pole in the scattering amplitude \cite{kappapoles1,kappapoles2,kappapoles3,JOPscatt}.
The most precise and model independent determination of its position in the second Riemann sheet was produced in Ref.~\cite{MoussallamKappa}.
Using Roy's equations for $K\pi$ scattering \cite{Roy} and Chiral Perturbation Theory
(ChPT)~\cite{ChPT} Descotes-Genon and Moussallam found
$m_\kappa = 658\pm 13$ MeV and
$\Gamma_\kappa = 557\pm 24$~MeV~\cite{MoussallamKappa}.

Although the experimental results are sound and the $\kappa$ pole is at present theoretically well known, a comprehensive and successful description of the reaction $\DKpp$ is still not available (for a recent review see Ref.~\cite{Reviews}).
Experimentalists, for the want of a better framework, commonly fit their data with the isobar model which consists of a weighted sum of Breit-Wigner-like propagators.
Often, a complex constant is added to the amplitude in order to account for the non-resonant decays.
It is known, nevertheless, that the adoption of Breit-Wigner functions to describe the effect of scalar resonances is problematic.

In the present work, we follow the general scheme where a factorised weak decay amplitude is dressed with FSIs by means of non-perturbative $K\pi$ form factors.
For the weak vertex, we employ the effective weak Hamiltonian of
Refs.~\cite{Wirbel:1985ji,Bauer:1986bm} within na\"ive factorisation.
The weak amplitude thus obtained receives contributions from colour-allowed and colour-suppressed topologies.
In the latter, the $K\pi$ form factors appear manifestly and the construction of the final state is straightforward.
The colour-allowed topology is more involved but, assuming the decay to be mediated by resonances as suggested by the experimental results, the FSIs in this case can also be written in terms of $K\pi$ form factors~\cite{Gardner:2001gc,Boito:2008zk}.
Therefore, in our description the hadronic FSIs are fully taken into account by the
$K\pi$ scalar and vector form factors.
Both form factors have received attention in recent years and are now well known in the energy regime relevant to $\DKpp$ decays.
The scalar component was studied in a framework that incorporates all the known theoretical constraints in Refs.~\cite{Jamin:2001zq,JOP2,JOP3}.
The results were subsequently updated and we employ in this work the state-of-the-art
version given in Ref.~\cite{Jamin:2006tj}.
The vector form factor, in its turn, can be studied in
$\tau^- \to K \pi \nu_\tau$ decays~\cite{Jamin:2006tk,JPP2,Moussallam,Boito:2008fq},
where the kinematical range is very similar to the one considered in this paper.
Here we employ a description which fulfils analyticity constraints and that was successfully fitted to the Belle spectrum in Ref.~\cite{Boito:2008fq}.

This contribution is based on a recent paper where all the issues discussed here are presented in more detail \cite{Boito:2009qd}.

\section{Theoretical framework}
\label{amplitudes}

Our phenomenological description of the process $D^+\to K^-\pi^+\pi^+$ is based on the
effective weak Hamiltonian.
At the quark level, the decay $D^+\to K^-\pi^+\pi^+$ is driven by the transition $c\to s u\bar d$,
\textit{i.e.}~four different quark flavours are involved.
In this case, only the two tree operators of the weak Hamiltonian have to be taken into account.
The amplitude for $D^+\to K^-\pi^+\pi^+$ is given by the matrix element
$\langle K^-\pi^+\pi^+|{\cal H}_{\rm eff}|D^+\rangle$.
We assume the factorisation approach to hold at leading order
(in $\Lambda_{\rm QCD}/m_c$ and $\alpha_s$) and as a consequence
the amplitude is written in terms of colour allowed and suppressed contributions,
${\cal A}_1$ and ${\cal A}_2$ respectively, as
\begin{equation}
\label{amplitude}
{\setlength\arraycolsep{2pt} 
\begin{array}{rcl}
{\cal A}(D^+\to K^-\pi^+\pi^+)&=&
\displaystyle
\frac{G_F}{\sqrt{2}}\cos^2\theta_C(a_1 {\cal A}_1+a_2 {\cal A}_2)
+(\pi^+_1\leftrightarrow\pi^+_2)\\[2ex]
&=&
\displaystyle
\frac{G_F}{\sqrt{2}}\cos^2\theta_C
[a_1\langle K^-\pi^+_1|\bar s\gamma^\mu(1-\gamma_5)c|D^+\rangle
        \langle\pi^+_2|\bar u\gamma_\mu(1-\gamma_5)d|0\rangle\\[2ex]
&&
+a_2\langle K^-\pi^+_1|\bar s\gamma^\mu(1-\gamma_5)d|0\rangle
        \langle\pi^+_2|\bar u\gamma_\mu(1-\gamma_5)c|D^+\rangle]
+(\pi^+_1\leftrightarrow\pi^+_2)\ ,
\end{array}
}
\end{equation}
where the last term accounts for the presence of two identical pions in the final state.
For the QCD factors $a_{1,2}$ we use the phenomenological values
$a_1=1.2\pm 0.1$ and $a_2=-0.5\pm 0.1$,
obtained from different analyses of two-body $D$ meson decays \cite{Buras:1994ij}.
The non-perturbative hadronic matrix elements in Eq.~(\ref{amplitude})
involve several Lorentz invariant form factors \cite{Boito:2009qd}.
The amplitude ${\cal A}_2$ reads
\begin{equation}
\label{A2}
{\setlength\arraycolsep{2pt}
\begin{array}{rcl}
{\cal A}_2
&=&
\left[m_{K\pi_2}^2-m_{\pi_1\pi_2}^2-
\displaystyle\frac{(m_K^2-m_\pi^2)(m_D^2-m_\pi^2)}{m_{K\pi_1}^2}\right]
F_+^{K\pi}(m_{K\pi_1}^2)F_+^{D\pi}(m_{K\pi_1}^2)\\[2ex]
&&
+\displaystyle\frac{(m_K^2-m_\pi^2)(m_D^2-m_\pi^2)}{m_{K\pi_1}^2}
F_0^{K\pi}(m_{K\pi_1}^2)F_0^{D\pi}(m_{K\pi_1}^2)\ ,
\end{array}
}
\end{equation}
where the Mandelstam variables are defined as
$m_{K\pi_1}^2\equiv (p_K+p_{\pi_1})^2$, 
$m_{K\pi_2}^2\equiv (p_K+p_{\pi_2})^2$ and
$m_{\pi_1\pi_2}^2\equiv (p_{\pi_1}+p_{\pi_2})^2$
with $m_{K\pi_1}^2+m_{K\pi_2}^2+m_{\pi_1\pi_2}^2=m_D^2+m_K^2+2m_\pi^2$.
Here, we follow Ref.~\cite{Gardner:2001gc}
and write the colour allowed amplitude ${\cal A}_1$ in terms of the scalar and vector $K\pi$ form factors as
\begin{equation}
\label{A1final}
{\cal A}_1=f_\pi\chi_S^{\rm eff}(m_D^2-m_{K\pi_1}^2)F_0^{K\pi}(m_{K\pi_1}^2)
+f_\pi\chi_V^{\rm eff}N(m_{K\pi_1}^2)F_+^{K\pi}(m_{K\pi_1}^2)\ ,
\end{equation}
where $\chi_S^{\rm eff}$ and $\chi_V^{\rm eff}$ two free parameters that are fixed from experimental branching ratios.

\section{Numerical results}
\label{results}
\begin{figure}[!ht]
\includegraphics[width=0.5\columnwidth,angle=0]{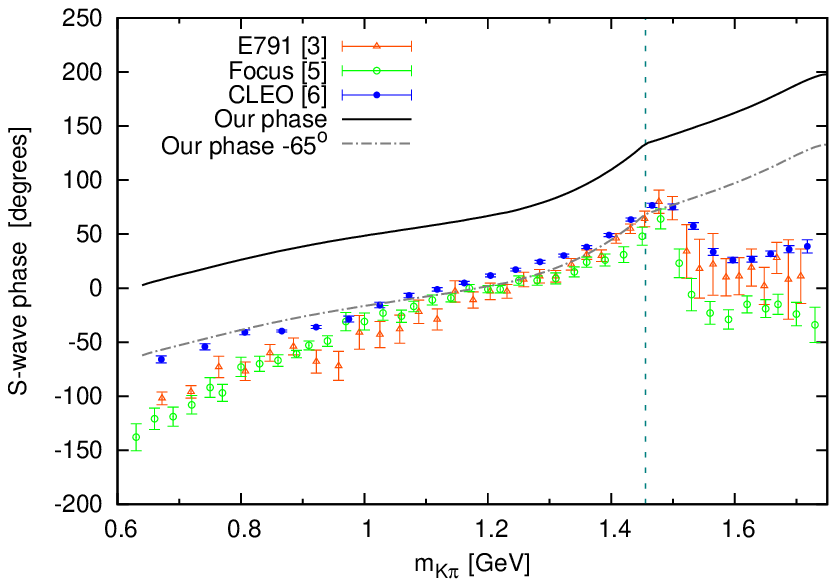}
\includegraphics[width=0.5\columnwidth,angle=0]{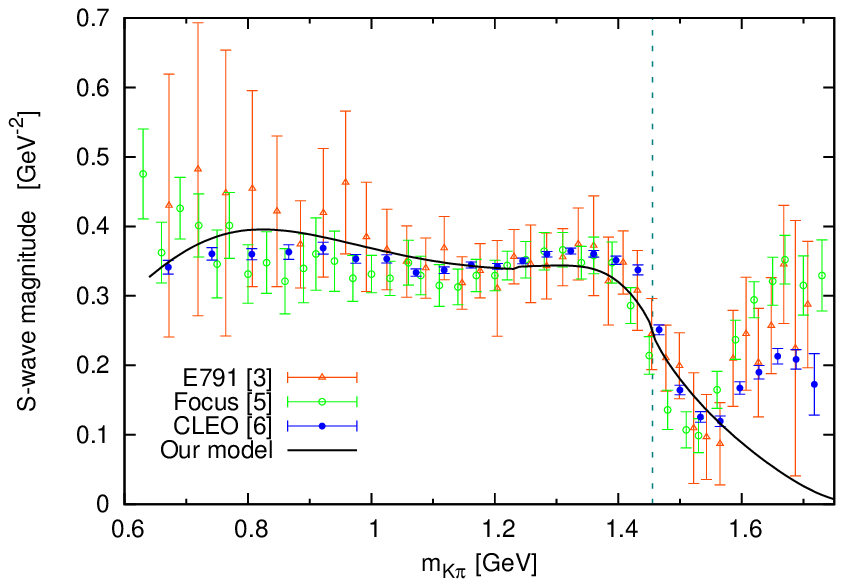}
\caption{{
(colour online).
(a) $S$-wave phases from the analyses of Refs.~\cite{Aitala:2005yh,Bonvicini:2008jw,Focus2009}. The solid line is the phase of our $S$-wave amplitude with $\alpha=0^\circ$,
whereas the dot-dashed line is  the $S$-wave phase  with $\alpha=-65^\circ$.
The dashed line delimits the $K\eta'$ threshold. 
(b) Absolute value of the $S$ wave  measured in
Refs.~\cite{Aitala:2005yh,Bonvicini:2008jw,Focus2009} compared with our model.
The amplitudes are normalised according to the text.}}
\label{Phase}
\end{figure}

Let us now investigate in detail the numerical results for our final model which includes the contribution of both $\A_1$ and $\A_2$ topologies.
The corresponding expressions are given in Eqs.~(\ref{A2}) and~(\ref{A1final}).
We begin by considering the $S$-wave description which is, in our opinion, the main aspect of the problem.
In our model, the $S$-wave FSIs are described by the $K\pi$ scalar form factor of
Ref.~\cite{Jamin:2006tj} in a quasi two-body approach,
{\it i.e.}, we assume that the $K\pi$ pairs in Eq.~(\ref{amplitude}) form an isolated system and do not interact with the bachelor pion.
Moreover, the form factor of Ref.~\cite{Jamin:2006tj} is obtained from dispersion relations that fix its phase to be the scattering one within the elastic region~\cite{Jamin:2001zq}.
Consequently, our $S$-wave amplitude has the $K\pi$ $I=1/2$ scattering phase up to roughly 1.45 GeV where the $K\eta'$ channel starts playing a role.
We compare in Fig.~\ref{Phase}a the experimental results from
Refs.~\cite{Aitala:2005yh,Bonvicini:2008jw,Focus2009} with the phase of our $S$ wave.
Since we are dealing with a production experiment, a global phase difference is expected as compared with scattering results~\cite{Pennington:2007se}.
Therefore, we allow for a global phase shift $\alpha$ in our $S$-wave amplitude
$\A_S(m_{K\pi_{1}}^2,m_{K\pi_{2}}^2)\rightarrow e^{i\alpha}\,
  \A_S(m_{K\pi_{1}}^2,m_{K\pi_{2}}^2)$.
In Fig.~\ref{Phase}a, we also plot as the dot-dashed line the phase of our amplitude shifted by
$\alpha=-65^\circ$.
With this shift, we see that up to 1.5 GeV CLEO's results and ours share a remarkably similar dependence on energy.
Inspired by the inspection of Fig.~\ref{Phase}a, we consider as our final model the one given by Eqs.~(\ref{A2}) and~(\ref{A1final}) with a shift of $\alpha=-65^\circ$ in the $S$-wave phase.

In order to compare the absolute value of our $S$ wave amplitude with experimental data,
we need fix the only two free parameters that occur in our model,
namely the normalisation constants $\chi_S^{\rm eff}$ and $\chi_V^{\rm eff}$.
The constant $\chi_V^{\rm eff}$ is fixed in order to reproduce the value of the sum of
all vector submodes.
Then, we fix the scalar normalisation $\chi_S^{\rm eff}$
requiring the total branching ratio from our model to match the world average.
Taking the central values for $a_1$ and $a_2$ this procedure gives
$\chi_S ^{\rm eff}= 4.9 \pm 0.4 \,\,\,\mbox{GeV}^{-1}$ and
$\chi_V^{\rm eff} = 4.4   \pm 0.6 \,\,\,\mbox{GeV}^{-1}$.
We can now compare the absolute value of our $S$-wave amplitude with experimental results.
However, since in isobar-like analyses the fit is sensitive only to the relative weights of the amplitudes, in order to compare the measurements with our result we need perform a normalisation.
We define a normalised $S$-wave amplitude by
$\A_S^{\mbox{\tiny Norm}}( m_{K\pi_{1}}^2, m_{K\pi_{2}}^2) =
 {\A_S}/{\left(\int_{\mathcal{D}} \,
 d  m_{K\pi_{1}}^2 d  m_{K\pi_{2}}^2\, | \mathcal{A}_S|^2\right)^{1/2}}$.
This amplitude, by construction, is  free of any global constants that appear in $\A_S$ and has dimension of [Energy]$^{-2}$.
Interpolating the results from the tables found in
Refs.~\cite{Aitala:2005yh,Bonvicini:2008jw,Focus2009}
we can calculate the normalised $S$ wave for each experiment.
We repeated the same procedure for our total $S$-wave amplitude.
The results for the $S$ wave are compared with our model in Fig.~\ref{Phase}b.

\begin{figure}[!ht]
\includegraphics[width=0.5\columnwidth,angle=0]{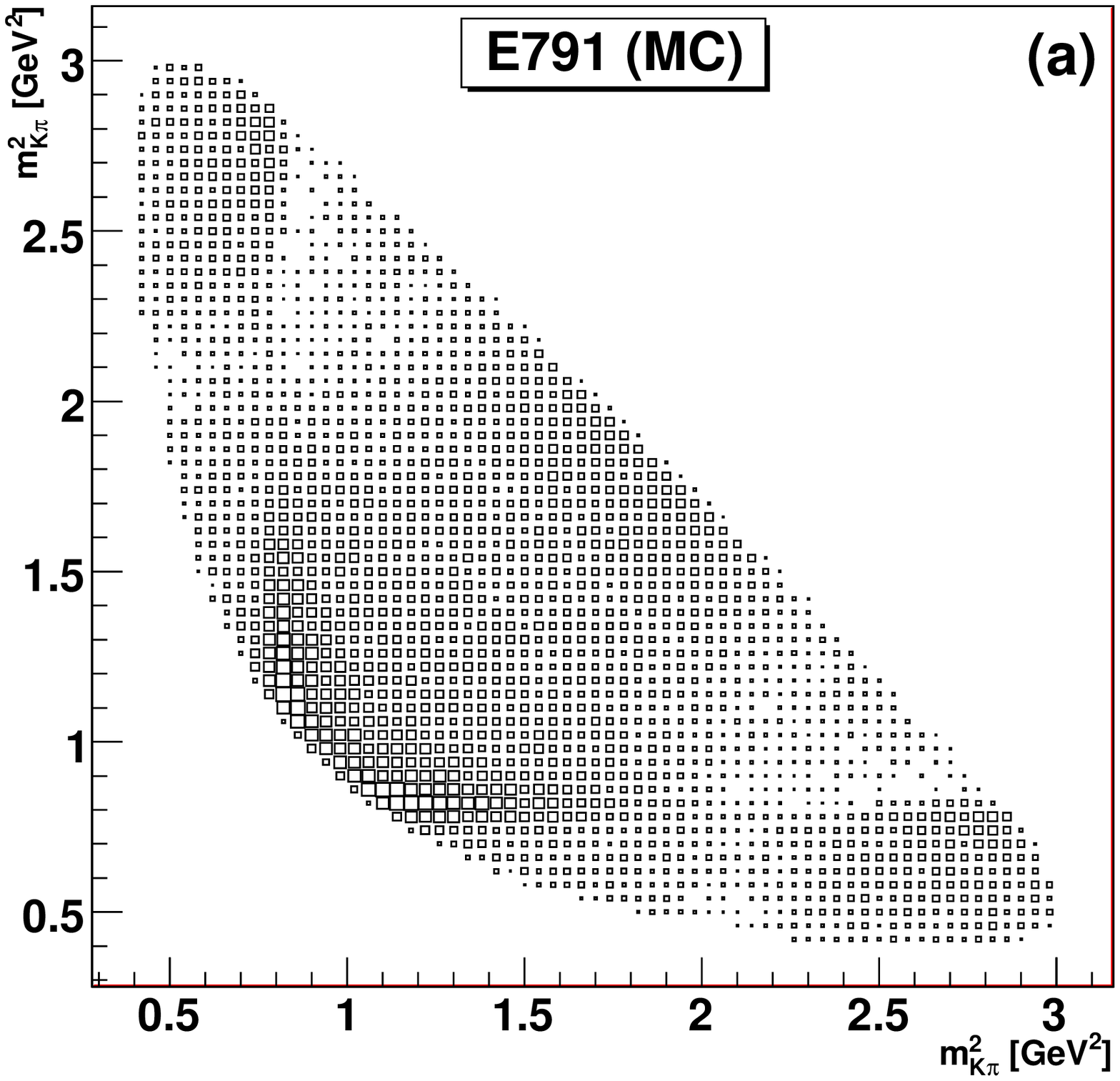}
\includegraphics[width=0.5\columnwidth,angle=0]{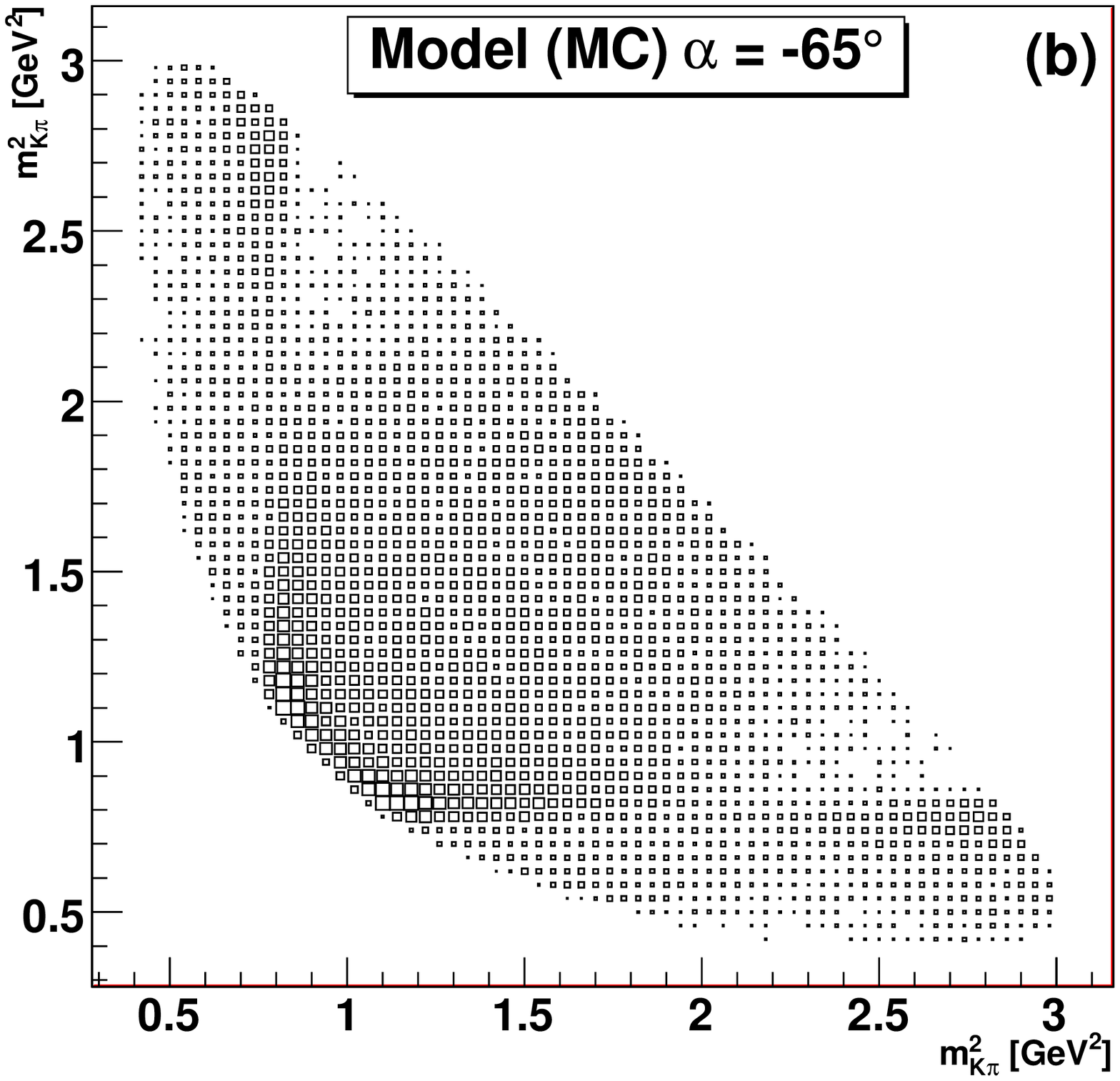}
\caption{{
(a) Monte Carlo simulation for the Dalitz plot of the E791 original analysis~\cite{Aitala:2002kr}
(b) Same for our model with a global shift of $-65^\circ$ degrees in the $S$-wave phase
(see text and Fig.~\ref{Phase}).
The number of independent events is 14185,
which correspond to the estimate of the signal events in Ref.~\cite{Aitala:2002kr}.}}
\label{Dalitz}
\end{figure}

\begin{figure}[!ht]
\includegraphics[width=0.33\columnwidth,angle=0]{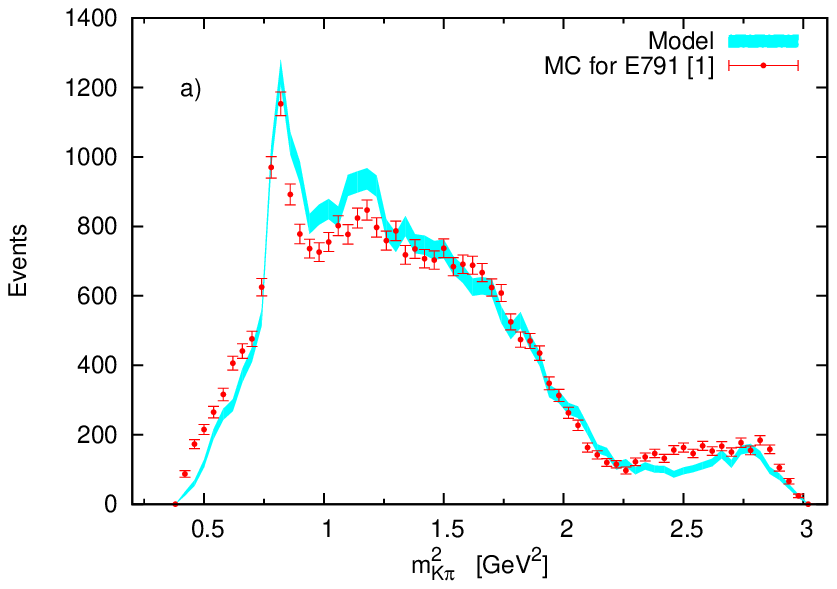}
\includegraphics[width=0.33\columnwidth,angle=0]{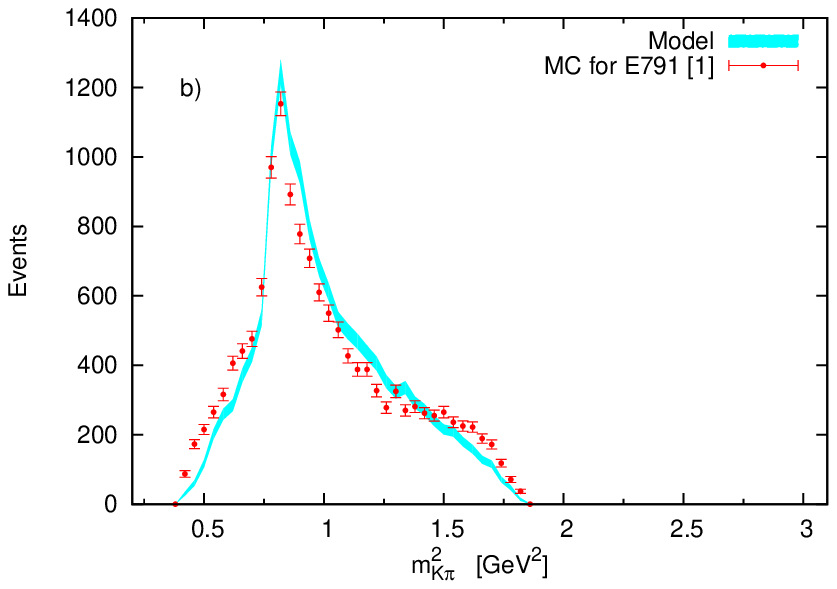}
\includegraphics[width=0.33\columnwidth,angle=0]{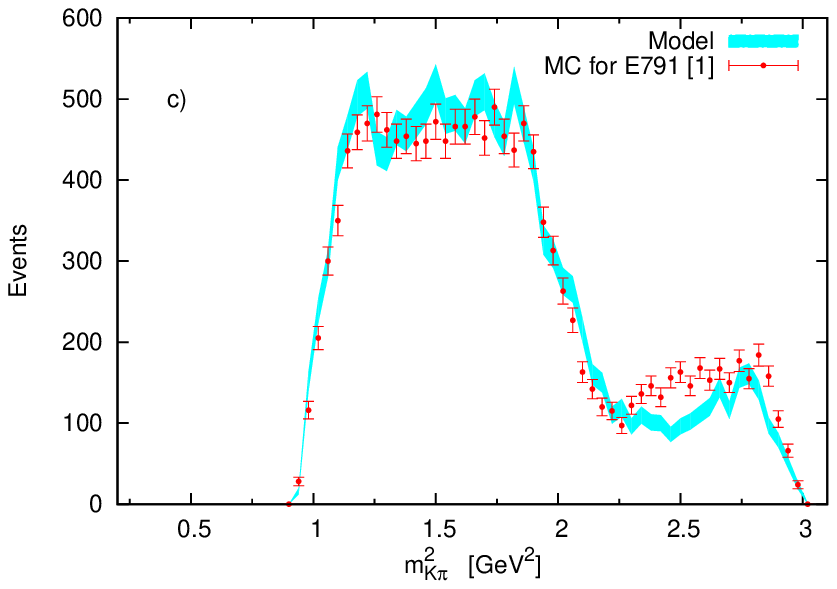}
\caption{{(colour online).
Projections from the MC generated Dalitz plots of Figs.~\ref{Dalitz}a and~\ref{Dalitz}b.
The error bars and the bands represent solely statistical fluctuations.
(a) Total projection, (b) high-energy projection, (c) low-energy projection.}}
\label{Projecs}
\end{figure}

We can also perform a Monte Carlo (MC) simulation to obtain a Dalitz plot from our model and compare the diagram and its projections with experimental results.
For the lack of a true data set,
we resort to a MC simulation of the original E791data~\cite{Aitala:2002kr}.
The obtained diagram is shown in Fig.~\ref{Dalitz}a.
Then we performed the same exercise for our model and the result is shown in Fig.~\ref{Dalitz}b.
Finally, in Fig.~\ref{Projecs} we show the projections of the diagrams of
Figs.~\ref{Dalitz}a and~\ref{Dalitz}b. 
 
\section{Summary and discussion}
\label{summary}
We have presented a model aimed at describing the decay $\DKpp$.
The weak amplitude is described within the effective Hamiltonian framework with the hypothesis of factorisation.
The $K\pi$ hadronic FSIs are treated in a quasi two-body approach by means of the well defined scalar and vector $K\pi$ form factors, thereby imposing  analyticity, unitarity and chiral symmetry constraints.
We used the experimental values for the total and $P$-wave branching ratios to fix the two free parameters in the model.
The relative global phase difference between the $S$ and $P$ waves was fixed phenomenologically using the experimental results of Ref.~\cite{Bonvicini:2008jw}.

The use of the $K\pi$ scalar form factor is shown to provide a good description of the $S$-wave FSIs.
Both the modulus and the phase of our $S$ wave compare well with experimental data up to
$m_{K\pi}\leq1.5$~GeV.
It is worth mentioning that the form factor we used has a pole that can be identified with the
$\kappa$.
Furthermore, the model is able to reproduce the experimental fit fractions and the total $S$-wave branching ratio.
Finally, the Dalitz plot arising from the model agrees with a MC simulated data set.

The main hypotheses of our model are the factorisation of the weak decay amplitude and the quasi two-body nature of the FSIs.
Therefore, the success of our description for $m_{K\pi}\leq 1.5$~GeV suggests that, in this domain, the physics of the decay is dominated by two-body $K\pi$ interactions.
We are led to conclude that effects not included in our model such as the $I=3/2$ non-resonant
$K\pi$ $S$ wave, the non-resonant $I=2$ $\pi^+\pi^+$ interactions and genuine three-body interactions, could be considered as corrections to the general picture described here.

Part of the discrepancy observed in our Dalitz plot is due to the disaccord of our $S$-wave amplitude for $m_{K\pi}\geq 1.5$~GeV.
A possible cause for this disagreement is the fact that factorisation in a three-body decay is expected to break down close to the edges of the Dalitz plot~\cite{BenoitBKpipi,Beneke}. Furthermore, in this region, the kinematical configuration of the final state momenta renders the quasi two-body treatment less trustworthy as well.
Finally, our model does not include the tensor component.
Although marginal, this amplitude has a non-trivial distribution in the phase space and could induce sizable interference effects in our plots.
In the vector channel, we find puzzling that the $K^*(1410)$,
which gives a sizable contribution for $\tau^-\to K \pi\nu_{\tau}$~\cite{JPP2,Boito:2008fq},
is hardly seen in experimental analyses of $\DKpp$.

In conclusion, since we do not fit the Dalitz plot we think that the agreement between the model and the experimental data is satisfactory.


\begin{theacknowledgments}
R.~E.~would like to express his gratitude to the HADRON 2009 Organizing 
Committee for the opportunity of presenting this contribution, and for the 
pleasant and interesting workshop we have enjoyed.
This work was supported in part by
the Ministerio de Ciencia e Innovaci\'on under grant CICYT-FEDER-FPA2008-01430, 
the EU Contract No.~MRTN-CT-2006-035482, ``FLAVIAnet'',
the Spanish Consolider-Ingenio 2010 Programme CPAN (CSD2007-00042), and
the Generalitat de Catalunya under grant SGR2009-00894.
\end{theacknowledgments}

\end{document}